\documentclass{mem}
\usepackage{natbib}\usepackage{txfonts}\usepackage{balance}
\usepackage{graphicx}
\usepackage[a4paper,breaklinks,dvipdfm]{hyperref}
\begin{document}
\def\teff{$T\rm_{eff }$}
\def\kms{$\mathrm {km s}^{-1}$}

\newcommand{\ba}{\begin{eqnarray}}
\newcommand{\ea}{\end{eqnarray}}
\newcommand{\be}{\begin{equation}}
\newcommand{\ee}{\end{equation}}
\def\eps{\epsilon}
\def\veps{\varepsilon}
\def\lesssim{\buildrel < \over {_{\sim}}}
\def\gtrsim{\buildrel > \over {_{\sim}}}

\title{
The blazar paradigm and its discontents
}

   \subtitle{}

\author{
C.\ D.\ Dermer }

\institute{
Code 7653, Naval Research Laboratory,
4555 Overlook Ave.\ SW, Washington, DC 20375 USA; 
\email{charles.dermer@nrl.navy.mil}
}

\authorrunning{Dermer }

\titlerunning{Blazar paradigm}

\abstract{
The standard blazar paradigm consists of a supermassive black hole that expels relativistic jets of magnetized plasma in our direction. This plasma entrains nonthermal synchrotron-emitting electrons that furthermore scatter internally-produced synchrotron photons as well as externally-produced photons from the accretion-disk, the broad-line region, and the infrared-emitting torus. This picture has been very successful in reproducing the two-humped blazar spectral energy distribution. Yet various discontents persist, including (1) ultra-short variability at TeV energies that is much shorter than the black hole's dynamical timescale; 
(2) very-high energy (VHE; $>100$ GeV) radiation from  FSRQs; (3) evidence for  a hard spectral component in high-synchrotron peaked objects, found when deabsorbing the measured VHE spectrum using conventional extragalactic background light (EBL) models; (4) an unusual slowly varying class of BL Lac objects; (5) contrary evidence about the location of the $\gamma$-ray emission regions.  Some of these problems can be resolved by introducing a hadronic component into the blazar paradigm, consistent with the hypothesis that blazars are sources of ultra-high energy cosmic rays. Gamma-ray observations with the Cherenkov Telescope Array and neutrino observations with IceCube will be important to test this hypothesis.
}
\maketitle{}


\section{Introduction}

%
Heber D.\ Curtis, who was on the right side of the famous Curtis-Shapely debate on the nature of spiral nebulae,  reported in 1918 that M87 displayed a ``\dots curious straight ray\dots apparently connected with the nucleus by a thin line of matter."  M87, which was discovered and cataloged even earlier by Charles \citet{mes1781}, remains one of the most interesting sources for the study of radio galaxies and blazars. At a distance of 16.4 Mpc, and with a black hole mass of $3.2(\pm 0.9)\times 10^9 M_\odot$ \citep{mac97},
Hubble Space Telescope observations of M87 show a one-sided jet with complex structure and optical knots traveling with apparent superluminal speeds as large as (4 -- 6)c \citep{bir99}, larger than radio superluminal speeds and large for an FR1 radio galaxy.  M87 also shows TeV variability on time scales of days \citep{aha06a}, a multiwavelength spectral energy distribution (SED) displaying a characteristic two-hump profile \citep{abd09a}, and has been implicated as the source of enhanced arrival directions of ultra-high energy cosmic rays (UHECRs) with energies $E \gtrsim 2\times 10^{19}$ eV \citep{sta95}.

The proximity of M87 and the small angle, $\sim 10^\circ$, of its jet to our line of sight, makes it a prototypical radio galaxy for the study of blazars. Typical of $\gamma$-ray emitting radio galaxies, it has strong core dominance \citep{abd10a} and displays properties in accord with schemes unifying FR1 radio galaxies and BL Lac objects \citep{up95}.

Interest in the study of radio galaxies and blazars has received a tremendous boost with the launch of the {\it Fermi} Gamma ray Space Telescope and the development of imaging ground-based $\gamma$-ray air-Cherenkov arrays, notably HESS, VERITAS, and MAGIC. Here we review the blazar paradigm and its successes and discontents, and some possible ways forward. The upcoming Cherenkov Telescope Array will be crucial for testing whether blazars are the sources of UHECRs.

\section{The blazar paradigm}

Blazars are active galaxies that exhibit rapid optical variability, strong radio and optical polarization, and superluminal motion. Based on extensive research, these observations are best understood in a paradigm where blazar emissions result from supermassive black holes powering relativistic jets that are pointed towards us. The apparent $\gamma$-ray luminosities of blazars averaged over time scales of years range from $\approx 10^{44}$ erg s$^{-1}$ to $\gtrsim 10^{49}$ erg s$^{-1}$ \citep{ack11}. During flaring episodes, the luminosity can be much higher. For example 3C 454.3 reached a record $\approx 2\times 10^{50}$ erg s$^{-1}$ \citep{abd11a} during an exceptionally active 5-day period  in 2010 November. The absolute luminosities are reduced from the {\it apparent} luminosities by a beaming factor $\sim 10^2$, which gives the fraction of the solid angle into which the radiation is beamed. Even so, the luminosities are enormous, considering that the Eddington luminosity for a black hole with mass $\approx 10^9 M_\odot$ is $\approx 10^{47}$ erg s$^{-1}$.  With measured variability times $t_{var}$ less than a day, and in many cases much shorter, causality arguments for stationary emission sites require that the radiation originates from a compact region of light days or less. The light-crossing time corresponding to the Schwarzschild radius of a $10^9M_\odot$ black hole is $\approx 10^4$ s. The simplest and only widely accepted explanation for the enormous luminosities and short timescales is that the blazar engine is a supermassive black hole. 

\subsection{Evidence for relativistic outflows}

Even a black-hole explanation is insufficient to account for the large powers, short variability times, and other unusual features of blazars. The emission region must furthermore be in bulk plasma expelled in a relativistic collimated outflow, representing an ``exhaust" valve for blazar engines quite distinct from those of radio-quiet AGNs which lack jets. Several lines of argument lead to this conclusion, including direct visual observations of jets in radio galaxies that are supposed to be misaligned blazars. The oldest such argument is the so-called {\it Compton catastrophe} \citep[e.g.,][]{jos74}. If the redshifts of blazars are really cosmological and their variable radio emission is nonthermal synchrotron from relativistic electrons, then the size scale implied by the variability time gives a radiation density so great that the radio-emitting electrons would Compton scatter the radio photons to high energies and make bright X-ray fluxes exceeding observations. Lacking such flux, it is important to note that the inferred internal radiation density is less for relativistic bulk plasma moving in our direction, so that the scattered X-ray flux will be less intense relative to the synchrotron emission. The second argument for relativistic jets is the phenomenon of {\it superluminal} motion, where the apparent transverse speed of radio-emitting blobs of plasma exceed $c$ \citep[e.g.][]{unw85}. As is well known, this is a kinematic effect resulting from radiating plasma moving at relativistic speeds $\beta c$ and bulk Lorentz factors $\Gamma$ at an angle $\theta \sim 1/\Gamma$ to our line of sight \citep[predicted by][]{ree66}. 

The third argument, and the one that is most relevant to $\gamma$-ray astronomy, is the strong $\gamma\gamma$ opacity implied if the emission region is at rest.  The EGRET instrument on the {\it Compton} Gamma ray Observatory showed that blazars are powerful sources of $\gtrsim 100$ MeV radiation \citep{har92}, but these $\gamma$ rays could never escape from the emission region due to strong pair production opacity from the process $\gamma + \gamma \rightarrow$ e$^+$ + e$^-$. The threshold for $\gamma\gamma$ pair production of a $\gamma$ ray with energy $\epsilon_1 = h\nu_1/m_ec^2$ interacting with ambient photons with energy $\epsilon = E_\gamma/m_ec^2$ is simply $\eps\eps_1 > 1$. Suppose that the emission region is at rest with inferred size $R\approx c t_{var}$ and photon luminosity $L_\gamma$, then the photon density of the target radiation field is $n_\gamma \approx L_\gamma/4\pi R^2 c E_\gamma$. The $\gamma\gamma$ opacity $\tau_{\gamma\gamma}\approx \sigma_{\gamma\gamma} n_\gamma R$, and $\sigma_{\gamma\gamma}\approx \sigma_{\rm T}$ for photons that exceed the threshold for pair production. The opacity, or compactness parameter, is therefore 
\begin{equation}
\tau_{\gamma\gamma} \approx {\sigma_{\rm T} L_\gamma\over 4\pi m_ec^4 t_{var}}\approx 10^3\, {L_\gamma/(10^{48}{\rm erg~s}^{-1})\over t_{var}({\rm d})}\,.
\label{taugammagamma}
\end{equation}
The assumption of a stationary emission region leads to the conclusion that $\gamma$ rays would be strongly attenuated, whereas blazars are powerful $\gamma$-ray sources.

The solution for these problems is bulk relativistic motion of the emission region. 
To illustrate, suppose that the emission region is in bulk relativistic motion, and can be described by a relativistic spherical shell of comoving width $\Delta R'$ and radius $R$. If the shell is briefly illuminated on a comoving time scale $\Delta t'\sim \Delta R'/c$, then due to strong beaming, an observer detects radiation that is emitted within the Doppler cone subtending an angle $\theta_{\rm Doppler}\approx 1/\Gamma$. Because of the time delay from off-axis emission regions, the radiation will 
vary to an observer on a time scale of $R(1-\cos\theta_{\rm Doppler})/c \approx R/2\Gamma^2 c$ (provided $\Delta R^\prime$ is sufficiently small), so that the emission region radius $R \lesssim 2\Gamma^2 c t_{var}$, which is $\sim \Gamma^2$ times larger than values inferred for a stationary emission region. In a shell geometry, the photon energy density in the comoving jet frame $u^\prime \sim L^\prime/4\pi R^2 \Gamma^2 c \sim L^\prime /4\pi c^3 t_{var}^2 \Gamma^6$.

By comparison in a blob geometry with observing angle $\theta$, the comoving luminosity $L_\gamma^\prime$ is Doppler boosted to the observer by four powers of the Doppler factor $\delta_{\rm D} \equiv [\Gamma(1-\beta\cos\theta)]^{-1}$, two powers for the reduction in solid angle into which the emission is directed, one power for the energy boost, and another power for the reduction in time over which the emission is received. Thus 
\begin{equation}
L_\gamma\approx \delta_{\rm D}^4 L^\prime_\gamma \approx 4\pi c R^{\prime 2}\delta_{\rm D}^4 u^\prime_\gamma \approx 4\pi c^3 \delta_{\rm D}^6 t_{var}^2 u^\prime_\gamma \;,
\label{Lgamma}
\end{equation}
where  the relation $R^\prime \approx c\delta_{\rm D} t_{var}$ characterizes the comoving size scale of the emission region (compare the similar expression for $u^\prime$ in a shell geometry). Consequently, the inferred photon energy density and number density are $\propto \delta_{\rm D}^{-6}$ and $\propto \delta_{\rm D}^{-5}$, respectively. This more dilute internal radiation field allows the $\gamma$ rays to escape without absorption. The condition $\tau_{\gamma\gamma}< 1$ requires only that $\Gamma \gtrsim 10$ in most blazars, though exceptional cases like PKS 2155-304 require $\delta_{\rm D} \gtrsim 60$ during an exceptional flaring state \citep{aha07a,bfr08}.

\subsection{Blazar SEDs}

Because of their extremely variable nature, multiwavelength campaigns have to be organized in order to obtain simultaneous or contemporaneous measurements of blazar SEDs. As a result of these campaigns, the multiwavelength spectra for perhaps a few dozen blazars are fairly well measured from radio to VHE (very high energy; $> 100$ GeV) $\gamma$-ray energies, excepting the $\approx 0.1$ -- 50 MeV regime where $\gamma$-ray telescopes still have relatively poor sensitivity. As has been apparent since the EGRET days, blazar SEDs exhibit a characteristic two-hump profile, where the lower frequency hump, which peaks in $\nu F_\nu$ from $\approx 10^{12.5}$ -- $10^{18}$ Hz,  is universally thought to be nonthermal synchrotron radiation from relativistic jet electrons. 

The higher frequency humps, which peak from $\lesssim 10$ MeV to VHE, are widely argued to originate from Compton scattering of ambient photons by the same electrons that radiate the synchrotron radiation.  The ambient synchrotron radiation provides a target field in all models. In synchrotron self-Compton (SSC) models, only the synchrotron photons provide targets. In external Compton (EC) models, photons are produced outside the jet and intercept the jet to be scattered. Models differ depending on the nature of the target photons for Compton scattering. For one-zone models, the principal external radiation fields in terms of energy density in the comoving jet frame \citep{sta09} are the direct accretion-disk field, the line radiation produced in the broad line region (BLR), the accretion-disk field scattered by electrons in the BLR, and the infrared radiation from a dusty torus surrounding the active nucleus \citep[see, e.g.,][for review]{2012rjag.book.....B}. In multi-zone models, additional radiation fields from different parts of the jet are considered, as in the decelerating jet model \citep{gk03} or the spine-layer model \citep{chi00}.

Different classes of blazars are defined according to various properties. BL Lac objects are typically defined if the equivalent width of the strongest optical emission line is $<5$\AA. By contrast, flat spectrum radio quasars (FSRQs) have strong optical emission lines indicating the presences of dense BLR material and strong illuminating accretion-disk radiation. 
Radio galaxies can be defined according to their radio luminosity and radio morphology \citep{fr74}. The twin-jet morphology of radio galaxies is seen in low-power radio galaxies, whereas the lobe and edge-brightened morphology is found in high-power radio galaxies, with a dividing line at $\approx 2\times 10^{25}$ W/Hz at 178 MHz, or at a bolometric radio luminosity of $\approx 2\times 10^{41}$ erg s$^{-1}$. Radio spectral hardness can also be used to characterize sources as flat spectrum, with number index harder than $1.5$, and steep spectrum, with softer indices. 
 
When there is sufficient multiwavelength coverage to reconstruct a spectrum from the radio through the optical and X-ray bands, blazars and radio galaxies can also be classified according to their broadband SEDs \citep{abd10b}. If the peak frequency $\nu^{\rm syn}_{pk}$ of the synchrotron component of the SED is $< 10^{14}$ Hz, then a source is called low synchrotron-peaked (LSP), whereas if the SED has $\nu^{\rm syn}_{pk}> 10^{15}$ Hz, then it is referred to as high synchrotron-peaked (HSP). Intermediate synchrotron-peaked (ISP) objects have $10^{14}$ Hz $<\nu^{\rm syn}_{pk}< 10^{15}$ Hz. Most FSRQs are LSP blazars, whereas BL Lac objects sample the LSP, ISP, and HSP range. According to the  unification scenario for radio-loud AGNs \citep{up95}, radio galaxies are misaligned blazars, and FR1 and FR2 radio galaxies are the parent populations of BL Lac objects and FSRQs, respectively. 

\subsection{Blazar spectral modeling}

The standard blazar paradigm consists of magnetized plasma ejected with relativistic speeds in a collimated outflow along the polar axes of a rotating black hole \citep[reviewed by][]{dm09,2012rjag.book.....B}. Nonthermal electrons are injected into the relativistic plasma and are assumed, for simplicity, to have an isotropic pitch-angle distribution in the comoving jet frame. The magnetic field $B^\prime$ in the jet frame is also assumed to be randomly oriented. The emission region is usually modeled as a comoving spherical emission region with radius $R^\prime$, though a spherical shell geometry produces similar results. Within this picture, the blazar SED is fit by assuming a form for the comoving electron distribution, choosing $\Gamma$ ($\delta_{\rm D}\cong \Gamma$ is usually assumed except when information about the jet angle is available), $B^\prime$, and $R^\prime$. The size of the emission region is related to the variability time through the relation $R^\prime \approx c \delta_{\rm D} t_{var}$.

\begin{figure}[]
\resizebox{\hsize}{!}{\includegraphics[clip=true]{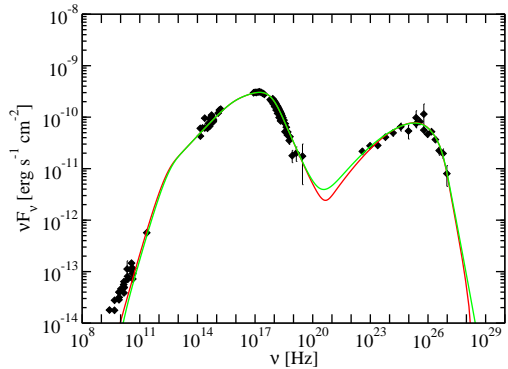}}
\caption{
\footnotesize
SED of Mrk 421 measured in multiwavelength campaigns including {\it Fermi} and {\it MAGIC} $\gamma$-ray telescopes \citep{abd11b}.  Two one-zone model fits are shown, with $t_{var} = 1$ hr and 1 day for the green and red curves, respectively.
}
\label{Mrk421}
\end{figure}

Considerable success has been achieved within this framework by fitting BL Lac objects with SSC models and FSRQs with EC models. For snapshot models, the electron distribution is chosen, whereas time-dependent models require information about the electron injection spectra, cooling, and adiabatic losses. Fig.\ 1 shows synchrotron/SSC modeling for the BL Lac object Mrk 421 \citep{abd11b}.
The electron distribution is described by a double broken power law, with $\delta_{\rm D} = 50$ and 21, and $B^\prime = 38$ and 82 mG for $t_{var} = 1$ hr and 1 day cases, respectively.  

\begin{figure}[]
\resizebox{\hsize}{!}{\includegraphics[clip=true]{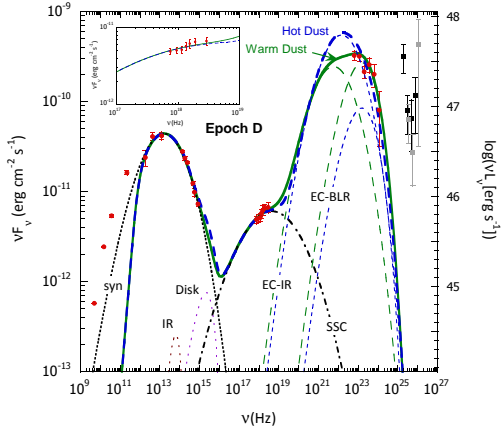}}
\caption{
\footnotesize
Blazar model fits \citep{der14} to the Epoch D SED of 3C 279
\citep{hay12}, showing the IR torus and accretion-disk 
emission, beamed synchrotron (syn), SSC, Compton-scattered IR (EC-IR)
radiaion for warm and hot dust, and Compton-scattered
BLR radiation (EC-BLR). Non-simultaneous VHE MAGIC data for 3C 279
are shown for comparison.  Total emission spectra include effects of 
synchrotron self-absorption. Inset shows detail of fits at X-ray energies.
}
\label{3C279}
\end{figure}

Fig.\ \ref{3C279} shows results of recent modeling efforts to fit the SED of 3C 279 \citep{hay12}. Rather than assigning values for $\delta_{\rm D}$ and $B^\prime$, equipartition was assumed between the nonthermal electron and magnetic field energy densities. Moreover, the electron distribution function is assumed to be described by a log-parabola function. Given these assumptions, we derive $\delta_{\rm D} =  22.5$ and $B^\prime = 0.41$ G for $t_{var} = 10^4$ s, and obtain a reasonably good fit for all the multiwavelength data excepting the VHE MAGIC data. 

Leptonic models have found considerable success in explaining multiwavelength blazar data. Futher examples of fits within the blazar paradigm are given for 3C 279 by  \citet{boe09}, for 3C 454.3 by \citet{bon11} and \citet{cer13a}, etc.

\subsection{The EBL}

The extragalactic background light, or EBL, is dominated by the background radiation from all the stars that have ever existed, either directly to make the optical-UV EBL, or through absorption and re-radiation by dust to make the IR EBL. The EBL is important because it contains integrated information about the cosmic evolution of matter and radiation through star formation, dust extinction, and light absorption and re-emission by dust. Knowledge of EBL absorption is needed to infer the intrinsic blazar spectra of extragalactic  sources of GeV and TeV $\gamma$ radiation that are primarily attenuated by optical and IR EBL photons, respectively, through the same process, $\gamma + \gamma \rightarrow$ e$^+$ + e$^-$, discussed earlier with respect to internal $\gamma$-ray absorption. 

The EBL is difficult to measure directly because of foreground zodiacal light and Galactic synchrotron radiation. Consequently, attempts are made to use $\gamma$-ray sources to infer the intensity of the EBL by examining its effects on blazar spectra. One approach is to argue from acceleration and radiation physics that the intrinsic blazar source spectrum cannot have a number spectral index harder than (i.e., greater than) $-1.5$ \citep{aha06b}. If the deabsorbed blazar VHE spectra is harder than this, the EBL model is ruled out. Another method \citep{geo10} is to extrapolate the Fermi spectrum of a blazar into the TeV range, and impose the condition that the deabsorbed VHE spectrum cannot exceed the extrapolation. A third method \citep{ack12} is to examine spectral cutoffs in the high-energy EBL spectra of a large sample of blazars at various redshifts $z$, and look for a systematic decrease in the $\gamma$-ray cutoff of the emission with increasing $z$. 

The effects of the EBL on $\gamma$ rays are represented by the value of $E_{\gamma\gamma}(z)$, which is defined as the measured photon energy emitted at redshift $z$ where the $\gamma\gamma$ opacity, $\tau_{\gamma\gamma}(E_{\gamma\gamma};z)$, equals unity. The value of $E_{\gamma\gamma}(z) \cong 1$ TeV for a source at  $z \cong 0.1$ \citep[e.g.,][]{fin10}. For Mrk 421 at $z = 0.03$ and 3C 279 at $z = 0.536$, $E_{\gamma\gamma} \cong 10$ TeV and 200 GeV, respectively. The EBL corrections to the models shown in Figs.\ 1 and 2 are negligible, though fits to the VHE radiation of 3C 454.3, and in BL Lac objects discussed below, definitely require EBL absorption corrections.

\section{Discontents in blazar studies}

Though the standard blazar paradigm has been very successful in accounting for the large luminosities and intense $\gamma$-ray fluxes from blazars, and the overall shape of their SEDs, persistent problems remain. We now describe some of these discontents.

\subsection{Rapid variability of BL Lac objects}

Blazars, by definition, are highly variable. Contrary to simple expectations that the light-crossing time across the Schwarzschild radius of the black hole that powers blazars sets a minimum variability time---which would be a few hours for a $10^9 M_\odot$ black hole---Mrk 421 at $z= 0.03$ \citep{for12}, Mrk 501 at $z = 0.033$ \citep{alb07} and PKS 2155-304 at $z = 0.116$ \citep{aha07a}  all display large amplitude variability on timescales of 10 minutes or less. This is more than an order of magnitude shorter than naively expected, and represents a feature of blazar physics that is not fully understood. 

\subsection{Unusual weakly variable BL Lac class}

In the opposite direction to the preceding item is the existence of a subclass of BL Lac objects that are unusually stable at TeV energies. These include 1ES 1101-232  at $z = 0.186$, 1ES 0347-121  at $z = 0.185$, and 1ES 0229+200 at $z = 0.14$. In the latter case, more than 8 separate {\it HESS} and {\it VERITAS} pointings over 7 years find $>300$ GeV fluxes that are all within 2$\sigma$ of the average flux \citep{cer13b}.

\begin{figure}[]
\resizebox{\hsize}{!}{\includegraphics[clip=true]{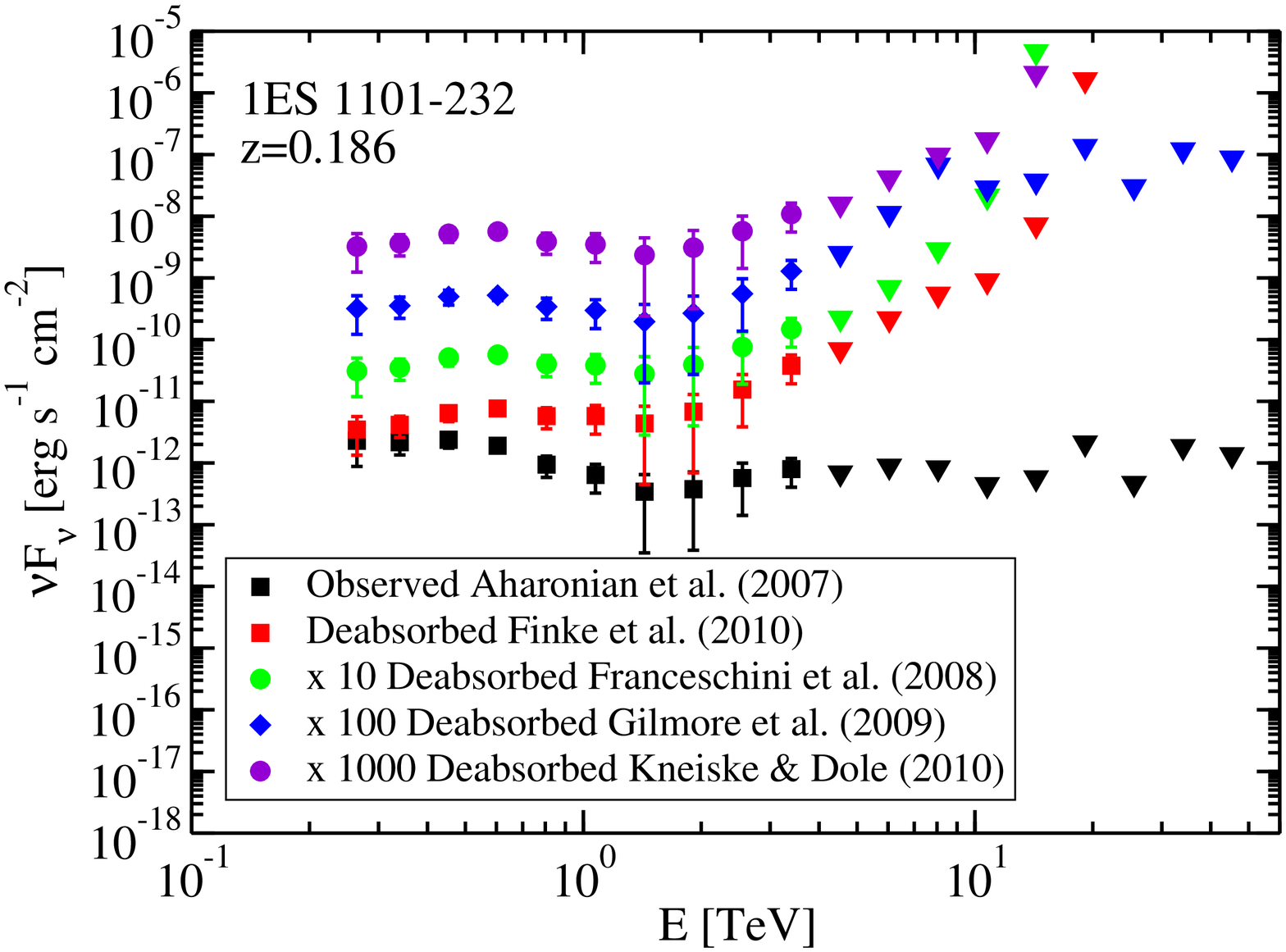}}
\caption{
\footnotesize
Measured $\gamma$-ray SED of 1ES 1101-232 \citep{aha07b}, compared with deabsorbed SEDs using different
EBL models, as noted in the legend.
}
\label{f3}
\end{figure}

\subsection{Hard blazar spectral components in deabsorbed VHE spectra}

The intrinsic blazar spectrum is expected to be obtained by multiplying the measured flux by the factor $\exp[\tau_{\gamma\gamma}(E;z)]$, where $\tau_{\gamma\gamma}(E;z)$ is given for a specific EBL model. When this is done, the deabsorbed flux reveals a hard spectral component in several cases (see Fig.\ 8 in \citet{fin10}). This is true for a wide range of EBL models, as can be seen in Fig.\ \ref{f3} for 1ES 1101-232. Should this spectral component be real, it would violate  assumptions used to infer the EBL intensity, namely that the VHE spectral index is softer than $-1.5$, or that the $\gamma$-ray SED monotonically declines with photon frequency.

\subsection{Flattening in the GeV - TeV spectral index difference with redshift}

More distant sources have their VHE emission increasingly attenuated with respect to nearby sources. Therefore the difference in the absolute values of the photon spectral indices at TeV and GeV energies should monotonically increase with $z$. Such a behavior was reported by \citet{ss06,ss10} for sources with $z\lesssim 0.1$, and by \citet{ek12} for higher redshift sources. Note that detailed analysis of the data taking into account energy bands for index measurement reduces the significance of this effect \citep{sfg13}.

\subsection{Conflicting results for the location of the $\gamma$-ray emission site in blazars}

From the discussion in Section 2.1, we saw that elementary kinematics indicates that $\gamma$-ray fluxes which vary on the time scale $t_{var}$ should be produced at a distance $R\lesssim 2\Gamma^2 c t_{var}/(1+z)$ from the central black hole (now including the redshift factor). For very short variability times, as in the case of 4C +21.35 with $t_{var} \cong 10$ m at VHE energies \citep{ale11a}, this implies that the emission region is well within the BLR for  typical values  of $\Gamma \approx 10$ -- 30 found in superluminal studies and spectral modeling of FSRQs. The VHE radiation would, however, be highly attenuated by $\gamma\gamma$ pair production if it originated inside the BLR \citep{dmt12}. Either $R \gtrsim 0.1$ pc, thereby requiring $\Gamma\gtrsim 100$, or the relationship between $t_{var}$ and $R$ is misunderstood (compare Section 3.1).

Correlations of high-energy $\gamma$-ray flaring events and times of radio and optical flares and polarization position angle swings leads to different conclusions about the location of the $\gamma$-ray production site. In the case of 3C 279, comparing Fermi-LAT $\gamma$-ray light curves with optical polarization and position angle directions suggests $\gamma$-ray emission in the pc-scale radio zone  \citep{abd10d}. By comparing radio, X-ray, and optical flux and polarization during a time when BL Lacertae was a TeV source,  \citet{mar08} argues that the $\gamma$-ray production occurs outside the radio core defined by synchrotron self-absorption. The GeV $\gamma$-ray emission site is argued to be $\sim 14$ pc away from the central engine in OJ 287 from multiwavelength light curve analysis of OJ 287 \citep{agu11}.

Of corresponding interest is the identification of $\gamma\gamma$ opacity features in blazar SEDs. \citet{ps10} and \citet{sp11} argue that the GeV break in LSP blazars could be due to pair production off He Ly$\alpha$ and He recombination radiation, requiring $\gamma$-ray production deep in the BLR. The availability of long baseline data sets of bright blazars with the Fermi-LAT allows the $\gamma$-ray blazar SEDs to be examined in some cases to $\sim 100$ GeV \citep[][see also \citet{sp14}]{brl14}. For a detailed treatment employing constraints arising from the intensity of the SSC component, see \citet{nbs14}.

\subsection{VHE ($> 100$ GeV) emission from distant FSRQs}

Most FSRQs have peak frequencies of the $\gamma$-ray $\nu F_\nu$ SED in the $\sim 10$ MeV -- 1 GeV range, and a break in the spectrum at a few GeV \citep{abd09b,abd10c}. The standard external Compton model for FSRQs can explain the break as due to Klein-Nishina effects when relativistic jet electrons scatter Ly $\alpha$ and other line photons of the BLR \citep[][see also Fig.\ \ref{3C279}]{cer13a}, or to combined scattering effects \citep{fd10}. Nevertheless, episodes of intense VHE emission have been recorded from 3C 279 \citep{alb08,ale11b}, PKS 1510-089 \citep{wag10,cor12}, and PKS 1222+216 \citep{ale11a}. A hard emission episode was also observed with the {\it Fermi} Large Area Telescope from 3C 454.3 \citep{pac14}.  Explanation of these features in leptonic scenarios requires unusual electron distributions far out of equilibrium.

\subsection{The Synchrotron Puzzle}

In Fermi acceleration scenarios, the acceleration timescale $t_{acc}$ should be much longer than the Larmor timescale $t_{\rm L}$, because a particle has to make several gyrations to accumulate a significant fraction of its energy. 
Equating the synchrotron energy-loss time scale with $t_{\rm L}$ implies a maximum synchrotron energy $\sim 100\Gamma$ MeV \citep[cf.][]{jh92}, many orders of magnitude greater than the peak or the maximum synchrotron frequency of blazars. The highest energy synchrotron photons from the Crab nebula or the delayed {\it Fermi} LAT emission from GRB 130427A \citep{ack14}  are in accord with or even exceed this value, so it is surprising that the maximum synchrotron energies of blazars is so much lower.


\section{Directions forward}

Current research on blazars is focusing on these puzzles. Some new directions in blazar research are now summarized.

\subsection{Acceleration physics}

Items 3.1 and 3.6 above suggest that our understanding of the acceleration mechanism in blazar jets is lacking. One approach to explain the short variability time scales is to consider direct electric-field acceleration that can circumvent limitations of the Fermi mechanism. In magnetic reconnection models \citep{gub10}, the short variability time is realized by an electron beam driven by reconnection taking place in a sub-volume of a larger region whose size scale is determined by the Schwarzschild radius, $R_{\rm Sch}$. To compensate for the small comoving size $R^\prime = f \Gamma R_{\rm Sch}$, with $f < 1$, reconnection is assumed to drive relativistic outflows or beams of particles with sufficiently large Lorentz factor that Doppler boosting compensates for the smaller available energy content in the small blob.  The observed luminosity  $L_{obs} \cong L_{iso} f^2 \left( {\delta_p/\delta_{\rm D}}\right)^4$ so, provided that the plasma Doppler factor $\delta_p$ and jet Doppler factor $\delta_{\rm D}$ are suitably chosen, the large apparent power is preserved even from the smaller size scale of the region. Other approaches include jet-within-jet or turbulent cell models \citep[e.g.,][]{mj10,np12,mar12}, and Poynting-dominated jet models \citep{nal12}. Plasma instability-induced short variability behavior has been considered by \citet{ssb12}.

\subsection{Hadronic models}

As discussed above, the deabsorbed SEDs of several BL Lac objects indicate the existence
 of a high-energy spectral component extending to TeV energies.
VHE emission in FSRQs is difficult to understand with leptonic models \citep{boe09}, 
so a new spectral component may also be required in this blazar class. A plausible 
explanation for these components is hadronic acceleration in blazar jets,
which can make VHE emission from photohadronic processes
\citep{ad03,boe09,boe13}.

From the \citet{hil84} condition, the maximum particle
energy is limited to energy $E <  \Gamma Ze B^\prime R^\prime$.
If UHECRs are accelerated in the inner jets of BL Lacs, 
then derived values of $\delta_{\rm D}$ and $B^\prime$ 
in SSC models for BL Lac objects show that protons can be accelerated 
to $\lesssim 10^{19}$ eV, and Fe nuclei to super-GZK energies \citep{mur12}. 
Accelerated hadrons that escape the jet will retain the 
collimation of the jet in which they were made.  If these
UHECR protons are not dispersed before entering intergalactic space \citep{mur12a}, 
energy dissipated through Bethe-Heitler pair production ($p +\gamma \rightarrow$ e$^+$ + e$^-$)
 during transit through the intergalactic medium while interacting with the 
 CMB and EBL can make a weakly or non-variable
$\gamma$-ray spectral feature  \citep{ek10,ess10,ek12}. This 
mechanism can account for the existence of a nonvariable VHE emission
component in BL Lac objects, and is furthermore consistent with the
Gpc distances of these weakly variable BL Lac objects which have
redshifts $z \approx 0.1$ -- 0.2 required to extract a significant fraction 
of energy of escaping UHECR protons through the Bethe-Heitler process.
The added emission from UHECRs in transit would also explain the alleged
flattening of the \citet{ss06,ss10} behavior of the spectral indices 
described  in Section 3.4 above \citep{ek12}.

Photohadronic production in FSRQs will make an escaping neutron beam that 
decays into protons \citep{ad03}. The escaping neutrons can deposit 
significant energy in FSRQs at the pc scale and beyond through photopion 
interactions with IR photons,
which will preserve the short timescale variability of the inner jet 
to explain emission and rapid variability 
of VHE emission at the multi-pc scale 
in 4C +21.35 \citep{dmt12}. 


UHECR production in FSRQs like 3C 279 might be revealed, as in the case of BL Lac objects,
by detection of a weakly variable cascade radiation induced
by photopion and photopair processes from beamed ultrarelativistic
protons traveling through intergalactic space.  Provided again that the UHECR
beam can escape from the structured regions surrounding 3C
279 without being dispersed, a slowly varying
UHECR-induced $\gamma$-ray halo should surround 3C 279. The
MAGIC detection of VHE emission from 3C 279 shows, however,
VHE emission that possibly varies on timescales less than
a day. Alternately, 
UHECRs can make a VHE contribution to the blazar SED from hadronic processes 
taking place in the jet. It remains to be seen whether modifications of 
leptonic scenarios or hadronic models with proton synchrotron or
photopion production and cascades \citep{boe13} are
preferred to make the VHE $\gamma$-ray spectra of FSRQs.

\begin{figure}[t]
\resizebox{\hsize}{!}{\includegraphics[clip=true]{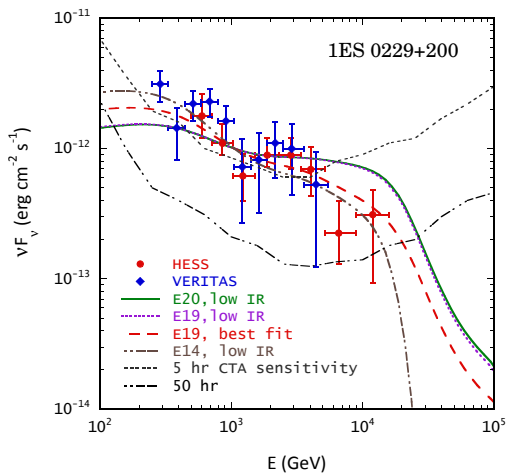}}
\caption{
\footnotesize
Spectral fits to HESS and VERITAS data of 1ES 0229+200.  
 The curves labeled ``E20, low IR" and ``E19, low IR" 
are the cascade spectra initiated by the injection of an UHECR proton distribution 
$\propto E^{-2}$  with $E_p^{\rm max} = {10}^{20}$ eV and ${10}^{19}$ eV protons, respectively, using a low-IR EBL model \citep[see][for details]{mur12}, whereas the curve labeled ``E19, best fit" is the spectrum with $E_p^{\rm max} = 10^{19}$ eV for the best-fit EBL model.  The curve labeled ``E14, low IR" is the spectrum resulting from the cascade of $E^{\rm max}={10}^{14}$ eV photons with hard source spectral index $=5/4$ for the low-IR EBL model.
}
\label{f4}
\end{figure}

\begin{figure}[]
\resizebox{\hsize}{!}{\includegraphics[clip=true]{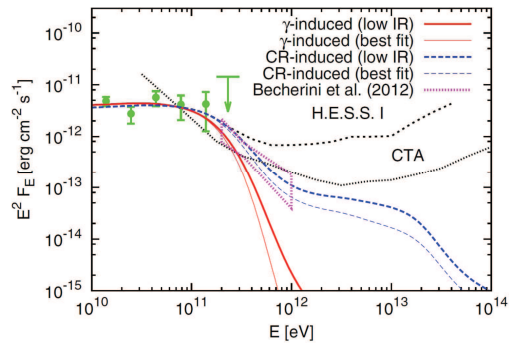}}
\caption{
\footnotesize
SEDs calculated for gamma-ray-induced (red) and UHECR-induced
(blue) cascade scenarios for KUV 00311-1938 ($z = 0.61$) using 
different EBL models \citep[see][]{tak13}.
}
\label{f5}
\end{figure}

The Cherenkov Telescope Array (CTA) could 
 solve the UHECR origin problem by distinguishing
between photon-induced and UHECR proton-induced cascades. 
Figs.\ 4 and 5 show these differences for the blazars 1ES 0229+200
and  KUV 00311-1938, respectively.
In Fig.\ 5, the analyzed LAT data (green) are shown 
with a preliminary HESS spectrum (magenta)
\citep{bec12}. 
The differential sensitivity curves 
for a 50 hr observation with HESS,
and the 5 and 50 hr sensitivity goals of the \citep[CTA configuration E;][]{cta10}, are shown. 
Note that the the HESS and CTA sensitivity criteria differ, and a
flux lower than the CTA sensitivity
curve can be achieved under a relaxed criterion of wider energy-bins and lower
significance for flux in each bin \citep[see][for details]{tak13}. The preliminary HESS data
is in agreement with the predicted SED made by UHECRs, but will require CTA to confirm.

\subsection{Neutrinos and new physics}

The most direct method to identify UHECRs in blazars is through detection 
of neutrinos. The IceCube detection of 28 and 37 neutrinos above the cosmic-ray induced
background \citep{aar13,aar14} has generated much excitement, though the number of neutrinos
is too small and directional uncertainties are too great to identify counterparts as yet
\citep[cf.][]{kra14}. \citet{pr14}
find associations of neutrino arrival directions with BL Lac objects and pulsar wind nebulae, and compare photon and neutrino SEDs of possible counterparts.
Both FSRQs \citep{mid13,dmi14} and BL Lac objects \citep{tav14}
have been argued to be the sources of the neutrinos. 

Solutions involving new physics, in particular, dark-matter particles, have
been proposed to resolve some problems in blazar physics. One example is the
axion, a light dark matter particle introduced to solve the strong CP problem in QCD.
Photon-axion conversion in the presence of a magnetic field can produce an
oscillation of photons to axion-like particles (and vice versa), leading 
to an enhancement of the received flux from a distant source. This mechanism
has been invoked to account for spectral features and GeV cutoffs in FSRQs \citep{tav12,mr13}.

\section{Conclusions} 

Blazar activity is robust both at the blazar sources and by the blazar scientists.
A standard paradigm has been accepted within which the overall properties and behaviors of 
blazars can be understood. Nevertheless, major puzzles have arisen that will require more 
study. Besides particle acceleration and cosmic rays, not even been mentioned in this short article
are open questions in blazar unification, sequence and divide, and the 
use of blazars for measurements of the magnetic 
field in the IGM  \citep[for recent reviews, see][]{ghi13,dn13}. Solving these problems and
obtaining a deeper grasp on reality, even if only the reality associated with
 supermassive
black holes with jets, should keep us busy for a long time.

\begin{acknowledgements}

I would like to thank Markus B\"ottcher, Piet Mientjes, 
and Brian Van Soelen for inviting me to South Africa and 
helping to support my visit. I would like to thank 
Justin Finke for the use of Fig.\ 3, and
Kohta Murase and Hajime Takami
for the use of Figs.\ 4 and 5. 
This work is partially supported by the Office of Naval Research.

\end{acknowledgements}

\bibliographystyle{aa}

\end{document}